\documentstyle[epsfig]{aipproc}
\def\dd{\delta}

\def\ga{\gamma}

\def\Th{\Theta}

\def\lra{\longrightarrow}
\def\gp{$\ga p \lra e^- e^+p$\/}
\def\ggg{$\ga \ga \lra e^- e^+$\/}
\begin{document}
\title{Collimated Energy-Momentum Extraction from Rotating
Black Holes\\in Quasars and Microquasars
Using the Penrose
Mechanism}

\author{Reva Kay Williams$^*$$^{\dagger}$
\footnote{E-mail: revak@astro.ufl.edu~~~~~~~~~~~~~~~~~~~~~~~~~
~~~\copyright  2001 The American Physical Society}}
\address{$^*$University of Florida, Gainesville, FL 32611\\
$^{\dagger}$Bennett College, Greensboro, NC 27401}

%\lefthead{LEFT head}
%\righthead{RIGHT head}
\maketitle

\begin{abstract}
For almost four decades, since the discovery of quasars, mounting
observational evidence has accumulated that black holes
indeed exist in nature. In this paper, I present a theoretical and 
numerical (Monte Carlo) fully relativistic 4-D analysis of
Penrose scattering processes (Compton and \ggg) in the 
ergosphere of a supermassive Kerr (rotating)
black hole.  These model calculations surprisingly reveal
that the observed high energies and luminosities of quasars and other 
active galactic nuclei (AGNs), the collimated jets about the polar
axis, and the asymmetrical jets (which can be enhanced by
relativistic Doppler beaming effects) all are inherent properties
of rotating black holes. From this analysis,  it is shown
that the Penrose scattered escaping relativistic particles
exhibit tightly wound coil-like cone distributions (highly
collimated vortical jet distributions) about the polar
axis, with helical polar angles of escape varying from $0.5^o$
to $30^o$ for the highest energy particles.  It is  also shown that 
the gravitomagnetic (GM) field, which causes the dragging of inertial 
frames, exerts a force acting on the momentum vectors of the incident 
and scattered particles, causing the particle emission to be asymmetrical 
above and below the equatorial plane, thus appearing to break the  
equatorial reflection symmetry of the Kerr metric. When the accretion disk 
is assumed to be a two-temperature bistable thin disk/ion corona 
(or torus $\equiv$ advection-dominated accretion flow),
energies as high as 54 GeV can be attained by these Penrose processes alone; 
and when relativistic beaming is included, energies in the TeV range can be 
achieved, agreeing with observations of some BL Lac objects.  When this 
model is applied specifically to quasars 3C 279 and 3C 273, their observed 
high energy luminosity spectra can be duplicated and explained. Moreover, 
this energy extraction model can be applied to any size black hole, 
irrespective of the mass, and therefore applies to microquasars as well.  
When applied specifically to microquasar GRS 1915+105 the results are 
consistent with observations.
\end{abstract}

\section{Introduction}

Astrophysical jets are one of the most poorly understood phenomena
today.
It is clear that they are present where gravitational accretion or 
contraction and magnetic
fields exist.  We observe these jets in quasars and microquasars due
to supermassive and stellar size black holes, respectively.  They
are also present in contracting protostars.  Perhaps, understanding 
the mechanism responsible for jets appearing from the energy source
of black holes, where the 
gravitational field is dominant, we can understand their appearences
associtated with photostars.  At present there are two 
popular trains of thought  associated with jets in black holes:
one is that the jets are inherent properties of geodesic 
trajectories in the Kerr \cite{kerr}
metric of a rotating black hole, and thus,
can be described by Einstein's general theory of relativity; and the other
is that the accretion disk and its magnetic field through 
magnetohydrodynamics (MHD) are producing the jets.  
Perhaps it could be a combination of the two,  with
gravity  controlling the flow near the event horizon,
and MHD controlling the flow at distances farther away.
In this paper, an anaylsis of the Penrose mechanism \cite{penrose}
is presented to 
describe gravitational-particle interactions close to the event horizon
in the subparsec regime.  In this fully general 
relativistic description, the jets are
produced and controlled by gravity alone, without
the necessity of an external magnetic field.

\section{Model}

The model consists of a supermassive $10^8 M_\odot$ Kerr
(rotating) black hole
plus particles from an
assumed
relativistic bistable thin disk/ion corona [or torus $\equiv$ 
advection-dominated accretion flow (ADAF)]:
two-temperature [separate
temperatures for protons ($\sim 10^{12}$~K) and
electrons ($\sim 10^9$~K)] 
 accretion
flow. 
 The Penrose mechanism
is used to extract rotational energy-momentum by
scattering processes inside the ergosphere
($r_{\rm erg}\simeq 2M$, where $c=G=1$).  
See Williams \cite{will1} for a detailed description of the
model. 
The
``quasi-Penrose'' processes investigated are
(a) Penrose Compton scattering (PCS) of equatorial
low energy
radially infalling
photons by equatorially confined ($Q_e=0$) and
nonequatorially confined ($Q_e\neq 0$) orbiting target electrons, 
at radii between
the marginally bound ($r_{\rm mb}\simeq 1.089M$) and marginally stable
($r_{\rm ms}\simeq 1.2M$) orbits;
(b) Penrose pair production (PPP) (\gp) at $r_{\rm mb}$;
(c) PPP (\ggg) by equatorial low energy
radially infalling
photons and high energy blueshifted (by factor $\simeq 52$)
nonequatorially
confined $\ga$-rays
at the {\it photon orbit} ($r_{\rm ph}\simeq 1.074M$).
 Note, the target particles are initially in bound (marginally stable
or unstable) trapped orbits, trapped in the sense of possibly having
no other way of escaping save these Penrose processes \cite{BPT,will1}.  
Note also that, as the nonequatorially
confined target particle passes through the equatorial plane,
$Q$, a constant of motion as measured by an observer at infinity
\cite{carter,will1}, equals $P_\Th^2$,
where $P_\Th$ is the polar coordinate momentum of the particle.

\section{Method}

Monte Carlo computer simulations of up to
$\sim 50,000$
scattering events of
infalling accretion disk photons (normalized to a
power-law distribution)  are executed for each quasar.
Energy and momentum (i.e., 4-momentum)
spectra of escaping particles
($\ga$'s, $e^-e^+$'s), as measured by an observer at infinity,
are obtained (where $\ga$ $\equiv$ a general photon, i.e., of any
energy value).
The following ingredients are used:
(1) General relativity [the Kerr metric spacetime
geometry yields  equatorially and nonequatorially confined
(``spherical-like'' \cite{wilk}) particle
orbits and escape conditions, conserved energy and angular momentum
parameters, and transformations from the Boyer-Lindquist
coordinate frame (BLF) to the local nonrotating frame (LNRF)].
Note, BLF is the observer at infinity \cite{bl};
LNRF is the local Minkowski (flat) spacetime.
(2) Special relativity [in the LNRF, physical
processes (i.e., the scatterings) are done; Lorentz transformations
between inertial frames are performed;
and Lorentz invariant laws are applied].
(3) Cross sections [application of the Monte Carlo method to the
cross sections, in the electron rest frame for PCS, in the proton
rest frame  for PPP(\gp), and in the center of momentum frame
for PPP(\ggg), give the  distributions of scattering angles and final
energies].

\section{Overall Results}

The energies attained are the following:
(1) {\it PCS}:
For the input (photon) energy range 5 eV to 1 MeV, the
 corresponding
output energy range is $\sim$ 15 keV to 14 MeV.
(2) {\it PPP} (\gp):
No escaping pairs for radially infalling
$\ga$-rays ($\sim$40 MeV), and no energy boost: implying that
the assumption (negligible recoil energy  given to the proton)
made in the conventional cross section and, perhaps,  
the geometry of the
scattering must
be modified.
It had been predicted \cite{lk}
 that pairs with
energies ($\sim $1 GeV) can escape.
(3) {\it PPP} (\ggg):
For input (photon) energy range $\sim$ 3.5 keV to 100 MeV, yields
output ($e^-e^+$) energy range $\sim$ 2 MeV to 10 GeV
(for BB),
and higher up to
$\sim$ 54 GeV (for PL, input $\ga \sim 2$~GeV), where
BB, PL$~\equiv~$blackbody and power-law distributions, respectively,
for the  accretion disk protons that yield the neutral pion decays
$\pi^0\lra\ga\ga$ \cite{eilek,MNK} to
populate the photon orbit.

The luminosity spectrum due to Penrose
processes for the specific case of quasar 3C~273 is plotted in
Figure~1(a), along with the observed spectrum
for comparison.
The outgoing (escaping) luminosity spectrum produced by the
Penrose scattered particles
is given by \cite{will2}
\begin{eqnarray*}
L_\nu^{\rm esc}&\approx&4\pi d^2 F_\nu^{\rm esc}~~
({\rm erg/s\,Hz})\\
                &\approx&4\pi d^2h\nu^{\rm esc}
                f_1 f_2 \cdots f_n\,(N_\nu^{\rm in}-N_\nu^{\rm cap}),
\end{eqnarray*}
where $d$ is the cosmological distance
of the black hole source;
$F_\nu^{\rm esc}$ is the flux of escaping photons;
$N_\nu^{\rm in}$ and $N_\nu^{\rm cap}$ are the emittance of
incoming and
captured photons, respectively;
$f_n$
defines the total fraction of the particles that undergoes scattering
[$n=2$ for PCS and $n=5$ for PPP (\ggg)].
The values of $f_1,\ldots, f_n$ are the fitting factors,
which make the Penrose calculated
luminosities agree with observations
for the specific case of 3C~273.
Note, in the model calculations, if we let every particle
scatter, and set $f_1=f_3\sim 10^{-2}$, defining the fraction of the
disk luminosity intersecting the scattering regime, with 
the remaining
$f_n$'s equal 1, the continuum emissions (the 
top curves on Figures~1(a), labeled
with numbers for specific cases of target and incident particles) are
obtained; see \cite{will2} for details.   The spectrum resulting from
the PPP (\ggg) is produced by letting the escaping pairs undergo 
``secondary Penrose Compton scattering'' with low energy 
radially infalling 
accretion disk photons ($\equiv f_3$).

Thus, as one can see from Figure~1(a), the Penrose mechanism
can generate the necessary luminosity observed, and the three
model calculated regions of emission [due to PCS by equatorial 
confined targets 
(curve passing through nos.~$1-7$),
by nonequatorially confined targets that cross  the 
equatorial plane (curve passing through nos.~$8-13$), and 
PPP (\ggg) (curve passing nos.~$14-25$)] are consistent with
the three major regions of emission in all quasars and AGNs.  
Note, the target photons  at the photon orbit can only exist 
in nonequatorially confined orbits \cite{will1}.
 
Moreover, the observed spectra of microquasars (or galactic black
holes), in general, appear not to have PCS emission by the nonequatorially 
confined target electrons, as well as the highest energy $\ga$-ray 
emission due to PPP (\ggg), indicating that these sources may not 
have an ion corona nor ADAF,  which 
would be need to populate the orbits to generate such emission.

\section{The Gravitomagnetic Field}

The gravitomagnetic (GM) force field is the gravitational analog
of a magnetic field.
It is the additional
gravitational force that a rotating mass
produces on a test particle.  The
GM force 
is produced by the gradient of
${\vec \beta_{_{\rm GM}}}= -\omega{\bf \hat e_\Phi}$,
where $\omega$ is the frame dragging velocity and
${\vec \beta_{_{\rm GM}}}$ is gravitomagnetic potential
\cite{TPM}.
Analysis of the equations governing the trajectories of the 
particles shows that 
the GM force, which acts
proportional to the momentum of a particle,  alters the
incoming and outgoing momentum parameters  of the incident 
and scattered particles, resulting in
asymmetrical distributions, thus, appearing to break
the reflection symmetry of the Kerr metric above and below the
equatorial plane \cite{will1,will3,will4}.  Effects
of the GM force acting on the PPP (\ggg) process can be discerned from 
Figures~1(d), 1(e), and~2.   Note, only the distribution for 
$(P_{\ga 2})_\Th>0$ is
shown in Figure~1.  When half of the 2000 target photons are allowed to
have  $(P_{\ga 2})_\Th>0$ and the other half  $(P_{\ga 2})_\Th<0$ 
of equal absolute values,  the $e^-e^+$ ``jet/counter-jet'' 
achieve a maximum ratio $\sim 3:1$, favoring $(P_\mp)_\Th>0$ 
\cite{will3}, as seen in Figure~2.
Polar coordinate momentum distributions, $(P_{\rm ph}^\prime)_\Th$, 
for PCS escaping photons are displayed in Figure~3.   Notice the 
effects of
the GM force field causing the photon jet/counter-jet 
to vary from nearly symmetric 
to asymmetric
for the different cases shown.  Of these cases the largest ratio 
achieved is $\sim 5:1$ [Fig.~3(c)].
The direct cause of the asymmetry in the polar direction appears
to be due to the severe frame dragging in the ergosphere in which
the GM field lines are dragged in the direction that the black hole
is rotating \cite{will3}.

\section{The Vortical Orbits Produced}

It is found that the Penrose scattered
particles escape along vortical trajectories 
collimated about the
polar axis \cite{will1,will2,will5}.
These distributions  are fluxes of coil-like
trajectories of relativistic jet particles escaping concentric
 the polar axis.
The highest energy escaping particles
have the largest $P_\Phi$ values, the near largest  $P_\Th$
values, and smallest $\vert P_r\vert$ values.  Note, $P_r$ is negative
(inward toward the polar axis) for most of the PCS photons, 
and positive 
for the $e^-e^+$ pairs.
The helical angles of escape  $(\dd_i)_{\rm esc}=
\vert 90^\circ -\theta\vert$ of
particle type $i$,
for the highest energy scattered
particles are $(\dd_{\rm ph})_{\rm esc}\simeq1^\circ -
30^\circ$ for PCS and
$(\dd_\mp)_{\rm esc}\sim 25^\circ- 0.5^\circ$
for the $e^-e^+$ pairs.  The above
characteristics of the escaping particles imply
 strong collimation about the polar axis,
giving rise to relativistic jets with
particle velocities up to
$\sim c$ [eg.~cf. Figs.~1(b), 1(c), 1(f), and~2].  Note, 
such vortical trajectories and 
collimation are consistent with the findings of de~Felice et~al.
 from spacetime geometrical studies of  general particle geodesics
in a Kerr metric \cite{dec1,dec2,dez}.

\section{Conclusions}

From this model to extract energy-momentum
from a black hole we can conclude the following:
PCS is an effective way to boost
soft x-rays to hard x-rays and  $\ga$-rays up to $\sim$ 14~MeV.
PPP (\ggg) is an effective way to produce
relativistic $e^-e^+$ pairs up to $\sim$ 54~GeV:
This is the probable mechanism  producing the fluxes of
relativistic pairs emerging from cores of AGNs.
These Penrose processes can operate for any size
rotating black hole, from quasars to
microquasars \cite{will6}.
Overall, the main features of quasars:
(a) high energy particles (x-rays, $e^-e^+$ pairs, $\ga$-rays)
coming from the central source;
(b) large luminosities;
(c) collimated jets;
(d) one-sided (or uneven jets),
can all be explained by these Penrose processes to
extract energy from a black hole.

Moreover, it is shown here that the geodesic treatment of 
individual particle processes
close to the event horizon in the subparsec regime,
as governed by the black hole,
is sufficient to described the motion of the particles.
This finding is consistent with MHD through the statement made by
 de Felice and Zanotti \cite{dez}, that
the behavior of such  individual particles on geometry 
(or gravity)-induced
collimated trajectories is also that of the
bulk of fluid elements in the guiding center approximation.
In light of this, MHD  should be incorporated
into these calculations to describe the flow of the
Penrose escaping particles away from the black hole, i.e.,
to perhaps further
collimate and accelerate these jet particles out to the
observed kpc distances.

\section*{acknowledgments}

I first thank God for  His thoughts 
 and for making this research possible.  
Next, I thank 
Dr.~Fernando de Felice and Dr.~Henry Kandrup for their helpful comments
and discussions.  Also, I thank Dr.~Roger Penrose for his continual
encouragement.  I am grateful to the late Dr.~Robert (Bob) Hjellming for
his helpful discussions and cherished collaboration.  Part of this 
work was done at the Aspen Center for Physics.  This work was supported 
in part by the National Science Foundation and Bennett College.

%\subsection*{Atmospheric Model}

%\paragraph*{High-$T_c$ calculations.}

%\cite{key1}.

%\begin{figure}[t!] % fig 1
\begin{figure}[tbh] % fig 1
\centerline{\epsfig{file=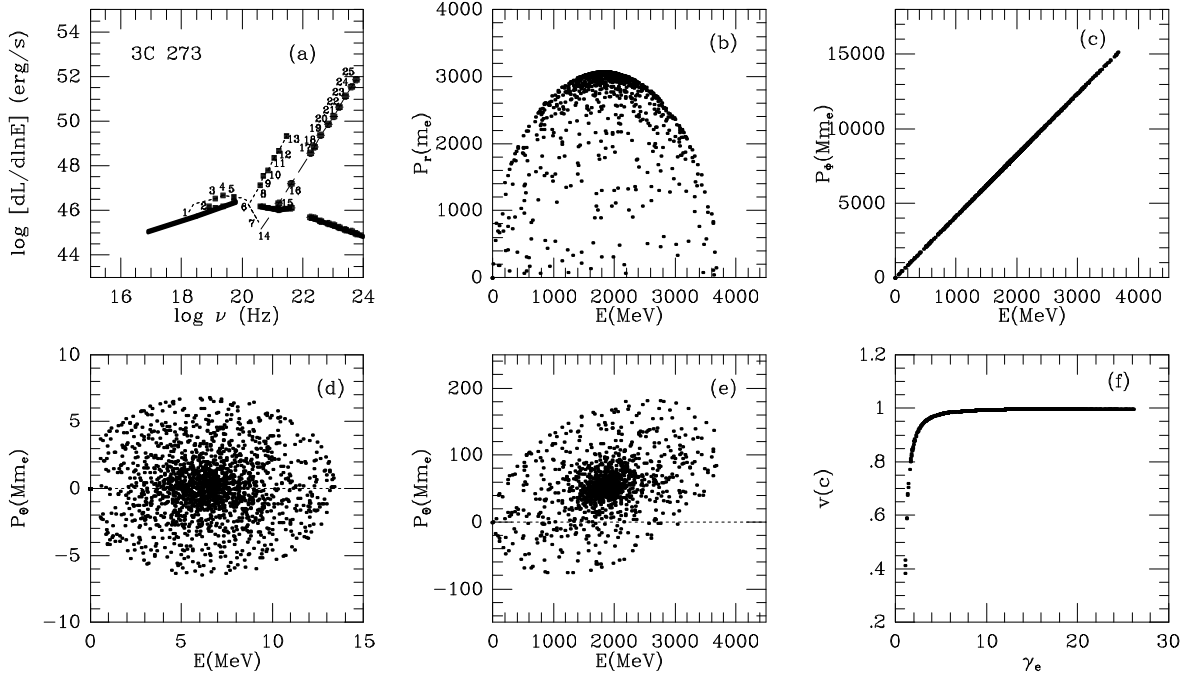,height=3.5in,width=4.5in,
bbllx=1.65in,bblly=1.in,bburx=4.5in,bbury=4.7in,angle=-90.}}
\vspace{35pt}
%\vskip 3.0in
\caption{(a) Comparing the theoretical spectrum
with observations for 3C 273. The calculated
PCS and PPP (\ggg) luminosity spectra are represented by the solid
squares and large solid dots, respectively.
The observed spectra is indicated
by the solid line.  The upper curves with the solid squares and
solid dots superimposed on the dotted line and the
dashed line, respectively,
for PCS and PPP (\ggg), are the spectra calculated from this model.
Superimposed on the lower solid line of the observations are
solid squares and solid dots that have been adjusted to agree with
observations.   These adjustments depend on the $f_n$'s values (see
text). (b) and (c) PPP
(\ggg) at $r_{\rm ph}=1.074M$:
scatter plots showing
momentum components (each point represents a scattering event).
The radial momenta $(P_\mp)_r$
vs.~$E_\mp$, the azimuthal
momenta $(P_\mp)_\Phi~(\equiv L_\mp)$ vs.~$E_\mp$;
for the infalling photons
$E_{\ga 1}=0.03$~MeV, and for the target photons
$E_{\ga 2}=3.893$~GeV.
(d) and (e) PPP (\ggg):  $(P_\mp)_\Th $ vs.~$E_\mp $;
for  $E_{\ga 1}=0.03$~MeV, $E_{\ga 2}=13.54$~MeV,
$(P_{\ga 2})_\Th=0.393\,Mm_e$;
 and for $E_{\ga 1}=0.03$~MeV,
$E_{\ga 2}=3.893$~GeV, $(P_{\ga 2})_\Th=113\,Mm_e$;
respectively. (f) The velocity distribution vs.
$\ga_e$($={E_\mp/ m_ec^2}$) for the same case as (d) above.
}
\label{fig1}
\end{figure}

%\begin{figure}[t!] % fig 2
\begin{figure}[tbh] % fig 2
\centerline{\epsfig{file=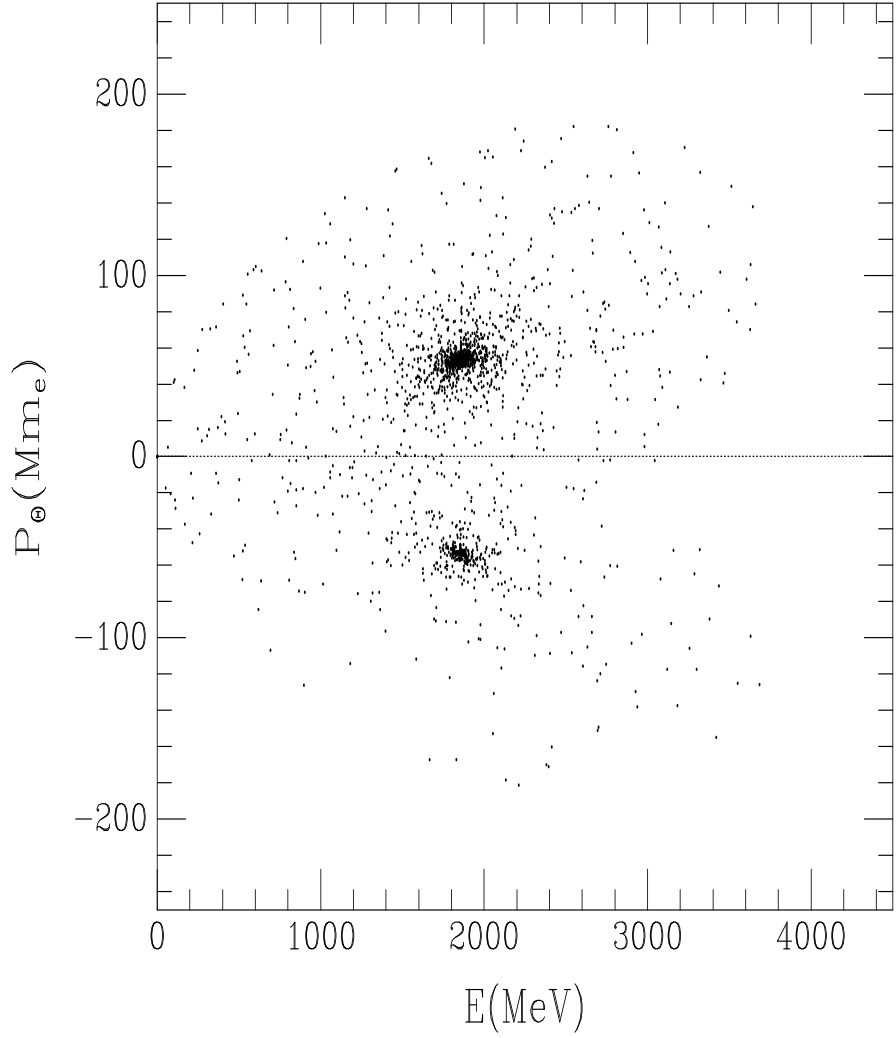,height=4.5in,width=3.5in,
bbllx=2.5in,bblly=2.in,bburx=7.0in,bbury=7.0in,angle=-90.}}
%\vspace{15pt}
\vskip 0.8in
\caption{PPP
(\ggg) at $r_{\rm ph}=1.074M$:
scatter plots showing
momentum components (each point represents a scattering event);
for the infalling photons
$E_{\ga 1}=0.03$~MeV, and for the target photons
$E_{\ga 2}=3.893$~GeV, $(P_{\ga 2})_\Th=\pm 113\,Mm_e$ [similar
to Fig.~1(e); see text].
}
\label{fig2}
\end{figure}

%\begin{figure}[t!] % fig 3
\begin{figure}[tbh] % fig 3
\centerline{\epsfig{file=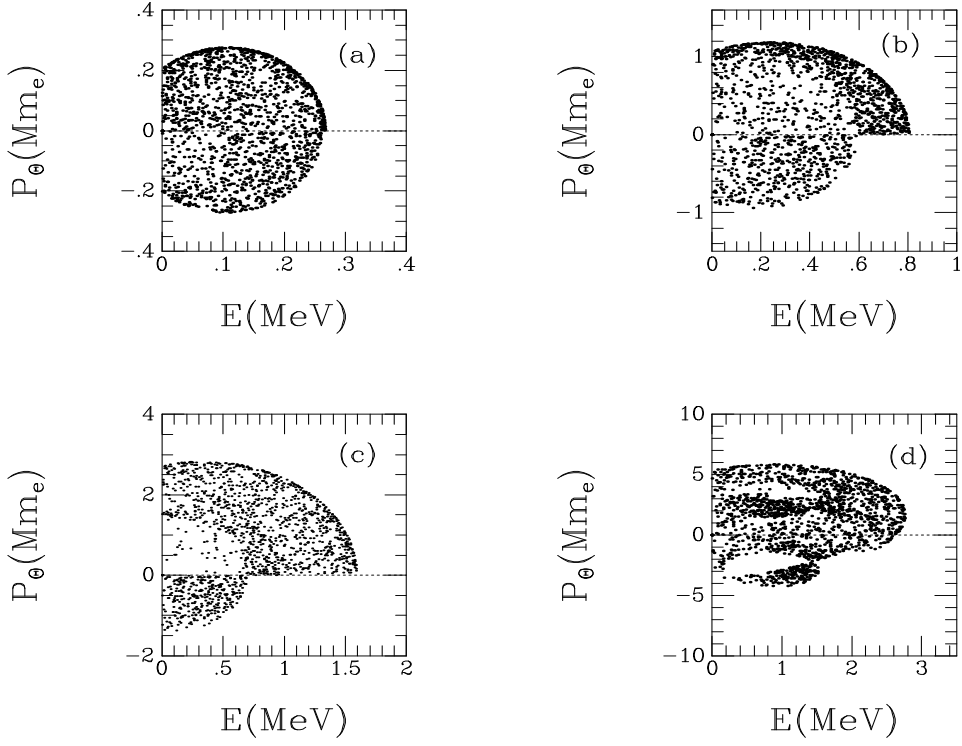,height=2.in,width=2.in,
bbllx=2.2in,bblly=6.0in,bburx=4.0in,bbury=8.0in,angle=0.
}}
%\vspace{10pt}
\vskip 0.5in
\caption{PCS: scatter plots showing
polar coordinate space momenta: $(P_{\rm ph}^\prime)_\Theta$
$[\equiv (Q_{\rm ph}^\prime)^{1/2}$]
vs.~$E_{\rm ph}^\prime$, of the escaping PCS photons after 2000 events
(each point represents a
scattering event), at $r_{\rm mb}\simeq 1.089M$.
The various cases are defined
by the following
parameters: $E_{\rm ph}$, initial photon energy; 
$E_e$, the target electron orbital energy;  $Q_{\rm e}^{1/2}$, 
defining the corresponding polar coordinate
momentum $(P_{e})_\Theta$ of the target electron;
$N_{\rm es}$, number
of photons escaping. (a) $E_{\rm ph}=3.5$~keV,
 $E_e=0.539$~MeV, $Q_{\rm e}^{1/2}=0$,
$N_{\rm es}=1637$.
(b) $E_{\rm ph}=0.03$~MeV, 
 $E_e=0.539$~MeV, $Q_{\rm e}^{1/2}=0$,
$N_{\rm es}=1521$.
(c) $E_{\rm ph}=0.15$~MeV, 
 $Q_{\rm e}^{1/2}=0$,
$N_{\rm es}=1442$.
(d) $E_{\rm ph}=0.15$~MeV,
 $E_e=1.297$~MeV, $Q_{\rm e}^{1/2}=\pm 2.479\,Mm_e$,
$N_{\rm es}=1628$. (Note, due to a minor
 oversight leading to improper treatment in the computer simulation
of the arccosine term in
 eq.~(3.39) of
[3], correct Figs.~3(a) and~3(b) presented here
replace Figs.~7(a)
and~3(c), respectively, of [3]. ) 
}
\label{fig3}
\end{figure}
\end{document}